\def\half{\frac{1}{2}}
\def\({\left (}
\def\){\right)}
\def\[{\left [}
\def\]{\right]}
\def\<{\left <}
\def\>{\right>}
\def\qed{\par\noindent\rightline{$\square$}}
\documentclass[a4paper,12pt]{article}
\usepackage{amssymb, amsmath}
\makeatletter
\renewcommand{\section}{{\setcounter{equation}{0}}\@startsection%
{section}%
{1}%
{0mm}%
{-\baselineskip}%
{0.5\baselineskip}%
{\normalfont\normalsize\bfseries}%
} \makeatother

\newcommand{\ds}{\displaystyle}
\newcommand{\ben}{\begin{enumerate}}
\newcommand{\een}{\end{enumerate}}
\newcommand{\be}{\begin{equation}}
\newcommand{\ee}{\end{equation}}
\newcommand{\bea}{\begin{eqnarray}}
\newcommand{\eea}{\end{eqnarray}}
\newcommand{\beas}{\begin{eqnarray*}}
\newcommand{\eeas}{\end{eqnarray*}}
\newcommand{\begth}{\begin{theorem}}
\newcommand{\enth}{\end{theorem}}
\newcommand{\blem}{\begin{lemma}}
\newcommand{\elem}{\end{lemma}}
\newcommand{\non}{\nonumber}
\newcommand{\nl}{\newline}
\newtheorem{theorem}{Theorem}[section]
\newtheorem{lemma}{Lemma}
\font\BB=msbm10
\def\RR{\hbox{\BB R}}

\def\CC{\hbox{\BB C}}

\def\trace{{\rm trace}\ }
\begin{document}
\markboth{ The Approximating Hamiltonian Method for the Imperfect Boson Gas}
{The Approximating Hamiltonian Method for the Imperfect Boson Gas}
{19-05-04}
\vskip3cm
\begin{center}
{\bf The Approximating Hamiltonian Method for the Imperfect Boson
Gas} \vskip 1cm {\bf Joseph V. Pul\'e} \footnote{{\it
Research Associate, School of Theoretical Physics, Dublin
Institute for Advanced Studies.}} \linebreak Department of
Mathematical Physics \linebreak University College
Dublin\\Belfield, Dublin 4, Ireland \linebreak Email:
Joe.Pule@ucd.ie \vskip 0.5cm and \vskip 0.5cm {\bf Valentin A.
Zagrebnov} \linebreak Universit\'e de la M\'editerran\'ee and
Centre de Physique Th\'eorique \linebreak CNRS-Luminy-Case 907
\linebreak 13288 Marseille, Cedex 09, France \linebreak Email:
zagrebnov@cpt.univ-mrs.fr
\end{center}
\vskip 2cm
\begin{abstract}
\vskip -0.7truecm

\mbox{}

\noindent The pressure for the Imperfect (Mean Field) Boson gas
can be derived in several ways. The aim of the present note is to
provide a new method based on  the Approximating Hamiltonian
argument which is extremely simple and very general.
\end{abstract}
\newpage
\section{Setup}
Consider a system of identical bosons of mass $m$
enclosed in a cube $\Lambda\subset \RR^d$ of volume V centered at
the origin. Let $E_0^\Lambda<E_1^\Lambda\leq E_2^\Lambda\leq
E_3^\Lambda\leq \ldots$ be the eigenvalues of
$h_\Lambda:=-\Delta/2m$ on $\Lambda$ with some boundary conditions
and let $\{\phi_l^\Lambda\}$ with $l=0,1,2,3,\ldots$ be the
corresponding eigenfunctions. Let $a_l:=a(\phi_l^\Lambda)$ and
$a^*_l:=a^*(\phi_l^\Lambda)$ be the boson annihilation and
creation operators on the Fock space ${\cal F}_\Lambda$,
satisfying $[a_l,a^*_{l'}]=\delta_{l,l'}$. Let $T_\Lambda$ be the
Hamiltonian of the free Bose gas, that is
$T_\Lambda=\sum_{l=0}^\infty E_l^\Lambda N_l$, where $
N_l=a^*_la_l $. Let $N_\Lambda=\sum_{l=0}^\infty N_l$ be the
operator corresponding to the number of particles in $\Lambda$.
The Hamiltonian of the interacting gas known as the
\textit{Imperfect} or \textit{Mean Field Boson Gas} is \be
H_\Lambda=T_\Lambda+\frac{a}{2V}N_\Lambda^2 \label{H} \ee where
$a$ is a positive coupling constant, (see e.g. \cite{HYL}).
\\
Let $\mu_0:= \lim_{\Lambda \uparrow  \RR^d}E_0^\Lambda$. Let $p_0(\mu)$
and $\rho_0(\mu)$ be the grand-canonical pressure and density respectively
for the free Bose gas at chemical potential $\mu<\mu_0$, that is:
\be
p_0(\mu)=-\int\ln(1-e^{-\beta(\eta-\mu)})F(d\eta) \ \ \ \ {\rm and}\ \ \ \
\rho_0(\mu)=\int\frac{1}{e^{\beta(\eta-\mu)}-1}F(d\eta),
\ee
$F$ being the integrated density of states of $h_\Lambda$ in the limit $\Lambda \uparrow
\RR^d$. Let $\rho_c:= \lim_{\mu\to \mu_0}\rho_0(\mu)$.
\\
The grand-canonical pressure of the Mean Field Boson Model with Hamiltonian (\ref{H}) is
\be
 p_\Lambda(\mu)=\frac{1}{\beta V} \ln \trace \exp\{-\beta (H_\Lambda-\mu N_\Lambda)\}
\ee and we put \be p(\mu)=\lim_{\Lambda \uparrow
\RR^d}p_\Lambda(\mu).\ee
{\bf Proposition.}\ \ {\it The pressure
in the thermodynamic limit $p(\mu)$ exists and is given by \be
p(\mu)=
\begin{cases}
\displaystyle{\half a\rho^2(\mu)+p_0(\mu-a\rho(\mu))} &{\rm if}\ \mu\leq \mu_c;\\
\\
\displaystyle{\frac{(\mu-\mu_0)^2}{2a} +p_0(\mu_0)} & {\rm if}\ \mu>\mu_c,
\end{cases}
\label{R} \ee where $\mu_c=\mu_0+a\rho_c$ and $\rho(\mu)$ is the
unique solution of the equation $\rho=\rho_0(\mu-a\rho)$.}
\par
This result, for special boundary conditions, can be proved in at
least three ways \cite{D, LdeSvdeB, DvdeBLP}, see also \cite{FV,
BP, ZP}. The aim of the present note is to provide yet another but
extremely simple and very general way of proving this result. Our
method covers also the case of \textit{attractive} boundary
conditions. The proof is based on the \textit{Approximating
Hamiltonian} technique, see e.g. \cite{BBZKT}.
\\
We shall need the following auxiliary operators for $\rho\in\RR$
and $\eta\in\CC$, \be H_\Lambda(\eta)=H_\Lambda +{\sqrt V}\(\eta
a^*_0 + \eta^* a_0\), \ee and \be
H_\Lambda(\rho,\eta)=T_\Lambda+a\rho N-\half a\rho^2 V+{\sqrt
V}\(\eta a^*_0 + \eta^* a_0\), \ee so that \be
H_\Lambda(\eta)-H_\Lambda(\rho,\eta)
=\frac{a}{2V}(N_\Lambda-V\rho)^2. \ee Let \be
p_\Lambda(\eta,\mu)=\frac{1}{\beta V} \ln \trace \exp \{-\beta
(H_\Lambda(\eta)-\mu N_\Lambda)\} \ee and \be
p_\Lambda(\rho,\eta,\mu)=\frac{1}{\beta V} \ln \trace \exp
\{-\beta(H_\Lambda(\rho,\eta)-\mu N_\Lambda)\}. \label{P2} \ee We
can write $H_\Lambda(\rho,\eta)-\mu N_\Lambda$ in the form \bea
H_\Lambda(\rho,\eta)-\mu N_\Lambda=\sum_{l=0}^\infty
\epsilon_l^\Lambda a_l^*a_l +{\sqrt V}\(\eta a^*_0 + \eta^*
a_0\)-\half a \rho^2 V \eea where \be
\epsilon_l^\Lambda(\rho,\mu):=E_l^\Lambda-\mu +a\rho. \ee For
convergence in (\ref{P2}) must have $a\rho -\mu +E_0^\Lambda>0$.
\par
\section{The Proof}
The proof of the Proposition consists of four
straightforward lemmas. The idea is to show that, for $\eta \neq 0
$, the pressure $p_\Lambda(\eta,\mu)$ in the limit coincides with
$p_\Lambda(\rho,\eta,\mu)$ minimized with respect to $\rho$
(Lemmas \ref{L0} and \ref{L1}) and that in turn this minimization
can be performed after the thermodynamic limit (Lemma \ref{L2}).
The final step is to switch off the source $\eta$ to obtain the
limiting pressure $p(\mu)$ (Lemma \ref{L3}). \blem \label{L0} For
a given $\eta$, there is a compact subset of $((\mu-\mu_0)/a,
\infty)$, independent of $\Lambda$, such that for $\Lambda$
sufficiently large the infimum of $p_\Lambda(\rho,\eta,\mu)$ with
respect to $\rho$ is attained in this set. \elem {\bf Proof:}\ \
From (\ref{P2}) we can see that \be
p_\Lambda(\rho,\eta,\mu)=-\frac{1}{\beta V}\sum_{l=0}^\infty\left
\{\ln\(1-\exp(-\beta \epsilon_l^\Lambda)\)\right \}
+\frac{|\eta|^2}{\epsilon^\Lambda_0} +\half a \rho^2 . \label{P}
\ee and \be \frac{\partial p_\Lambda}{\partial
\rho}(\rho,\eta,\mu) =-\frac{a}{V}\sum_{l=0}^\infty
\frac{1}{\exp(\beta
\epsilon_l^\Lambda)-1}-\frac{a|\eta|^2}{(\epsilon^\Lambda_0)^2}+a\rho.
\label{M} \ee Suppose $\eta\neq 0$. If  $a\rho<\mu-\mu_0 +\delta
$, with $\delta >0$, then \be \frac{\partial p_\Lambda}{\partial
\rho}(\rho,\eta,\mu)\leq
-\frac{a|\eta|^2}{(\epsilon^\Lambda_0)^2}+a\rho\leq
-\frac{a|\eta|^2}{(2\delta)^2}+\mu -\mu_0+\delta<0 \ee if $\delta$
is sufficiently small and $\Lambda$ large enough. On the other
hand for $a\rho >\mu-\mu_0 +\alpha $, with $\alpha >0$, \be
\frac{\partial p_\Lambda}{\partial \rho}(\rho,\eta,\mu)
>-\frac{a}{V}\sum_{l=0}^\infty \frac{1}{\exp(\beta (E^\Lambda_l-(\mu_0-\alpha))-1}
-\frac{4a|\eta|^2}{\alpha^2}+a\rho> -a(\rho_c-1)-\frac{4a|\eta|^2}{\alpha^2}+a\rho>0
\ee
for $\rho$ and $\Lambda$ sufficiently large. Therefore there exists $K<\infty$,
independent of $\Lambda$ such that the
infimum of $p_\Lambda(\rho,\eta,\mu)$ with respect to $\rho$ is attained in
$[(\mu-\mu_0+\delta)/a, K]$.
\qed
\par
Suppose the infimum of $p_\Lambda(\rho,\eta,\mu)$ with respect to $\rho$ that
is attained at ${\bar \rho}_\Lambda(\eta,\mu)$,
which is not {\it a priori} unique.

\blem \label{L1} If $\eta\neq 0$ \be \lim_{\Lambda \uparrow
\RR^d}p_\Lambda(\eta,\mu)=\lim_{\Lambda \uparrow
\RR^d}p_\Lambda({\bar \rho}_\Lambda,\eta,\mu). \ee \elem {\bf
Proof:}\ \ Since by Lemma \ref{L0}, ${\bar \rho}_\Lambda$ is a
interior point of $((\mu-\mu_0)/a, \infty)$, it satisfies \be
\frac{\partial p_\Lambda}{\partial \rho}({\bar
\rho}_\Lambda,\eta,\mu)=0. \label{N} \ee Now \be
\<\frac{N_\Lambda}{V}\>_{H_\Lambda(\rho,\eta)}=\frac{\partial
p_\Lambda}{\partial \mu}(\rho,\eta) =\frac{1}{V}\sum_{l=0}^\infty
\frac{1}{\exp(\beta
\epsilon_l^\Lambda)-1}+\frac{|\eta|^2}{(\epsilon^\Lambda_0)^2}.
\label{O} \ee Comparing this last equation (\ref{O}) with
(\ref{M}) we see that ${\bar \rho}_\Lambda$ satisfies the equation
\nl $\ds{{\bar \rho}_\Lambda
=\frac{1}{V}\<N_\Lambda\>_{H_\Lambda({\bar \rho}_\Lambda,\eta)}}$.
By Bogoliubov's convexity inequality, see e.g. \cite{BBZKT}, \be
0\leq p_\Lambda({\bar \rho}_\Lambda,
\eta,\mu)-p_\Lambda(\eta,\mu)\leq
\frac{1}{2V^2}\Delta_\Lambda(\eta). \ee where \be \Delta_\Lambda
(\eta)=a\<(N_\Lambda-V{\bar \rho}_\Lambda)^2\>_{H_\Lambda({\bar
\rho}_\Lambda,\eta)}. \ee We want to obtain an estimate for $
\Delta_\Lambda (\eta)$ in terms of $V$. Since \be
\frac{\Delta_\Lambda (\eta)}{aV}=\frac{\partial^2
p_\Lambda}{\partial \mu^2}(\rho,\eta)
=\frac{\beta }{V}\sum_{l=0}^\infty\frac{\exp(\beta \epsilon_l^\Lambda)}
{\(\exp(\beta\epsilon_l^\Lambda)-1\)^2}+\frac{2|\eta|^2}{(\epsilon_l^\Lambda)^3},\non \\
\ee
we use  $e^x/(e^x-1)\leq 2(1+1/x) $ for $x\geq 0$  and $\epsilon_l^\Lambda({\bar \rho}_\Lambda,\mu)
\geq E_0^\Lambda+a{\bar \rho}_\Lambda-\mu>E_0^\Lambda-\mu_0+\delta>\half \delta $ to obtain
\be
\frac{\partial^2 p_\Lambda}{\partial \mu^2}({\bar \rho}_\Lambda,\eta,\mu)\leq
2\(\frac{2}{\delta}+\beta\)\frac{1}{V}\sum_{l=0}^\infty\frac{1}{\(\exp(\beta
\epsilon_l^\Lambda)-1\)}+\frac{4}{\delta}\frac{|\eta|^2}{(\epsilon^\Lambda_0)^2}.
\ee
By (\ref{M}) and (\ref{N})
\be
\frac{1}{V}\sum_{l=0}^\infty\frac{1}{\(\exp(\beta \epsilon_l^\Lambda)-1\)}\leq{\bar \rho}_\Lambda
\ee
and
\be
\frac{|\eta|^2}{(\epsilon^\Lambda_0)^2}\leq{\bar \rho}_\Lambda.
\ee
Therefore
\be
\frac{\partial^2 p_\Lambda}{\partial \mu^2}({\bar \rho}_\Lambda,\eta,\mu)\leq
2\(\frac{4}{\delta}+\beta\){\bar \rho}_\Lambda.
\ee
Thus since by Lemma \ref{L0}, ${\bar \rho}_\Lambda<K$,
\be
\lim_{V\to\infty}\frac{1}{V^2}\Delta_\Lambda(\eta)=\lim_{V\to\infty}\frac{a}
{V}\frac{\partial^2 p_\Lambda}{\partial \mu^2}({\bar \rho}_\Lambda,\eta,\mu)=0.
\ee
\par
\qed
\par
\blem
\label{L2}
\be
\lim_{\Lambda \uparrow  \RR^d}p_\Lambda(\eta,\mu)=\half a\rho^2(\eta,\mu)+p_0(\mu-a\rho(\eta,\mu))
+\frac{|\eta|^2}{a\rho(\eta,\mu)-\mu}
\ee
where $\rho(\eta,\mu)$ is the unique solution of the equation $ \ds{\rho=\rho_0(\mu-a\rho)
+\frac{|\eta|^2}{(a\rho-\mu)^2}}$.
\elem
{\bf Proof:}\ \
Let $\rho(\eta,\mu)$ be a limit point of ${\bar \rho}_\Lambda\in[(\mu-\mu_0+\delta)/a, K]$.
$ p_0(\mu-a\rho)$ and $\rho_0(\mu-a\rho)$ are convex in $\rho$. From (\ref{P})
one can see that $p_\Lambda(\rho,\eta,\mu)$ is convex and thus as $\Lambda\uparrow\RR^d $,
$p_\Lambda(\rho,\eta,\mu)$
converges uniformly in $\rho$ on compact subsets of $((\mu-\mu_0)/a,\infty)$ to
\be
\half a\rho^2+p_0(\mu-a\rho)+\frac{|\eta|^2}{a\rho-\mu}.
\ee
Similarly from(\ref{M}),
one can see that
$\ds{\frac{\partial p_\Lambda}{\partial \rho}(\rho,\eta,\mu)+\frac{a|\eta|^2}
{(\epsilon^\Lambda_0)^2}} $
is also convex and thus by the same argument,
it converges uniformly in $\rho$ on compact subsets of $((\mu-\mu_0)/a,\infty)$.
Since it is also clear that $\ds{\frac{a|\eta|^2}{(\epsilon^\Lambda_0)^2}}$
converges uniformly on the same subsets, so does
$\ds{\frac{\partial p_\Lambda}{\partial \rho}(\rho,\eta,\mu)} $
with limit
\be
a\rho-a\rho_0(\mu-a\rho)-\frac{a|\eta|^2}{(a\rho-\mu)^2}.
\ee
Since
\be
\lim_{\Lambda \uparrow  \RR^d}p_\Lambda(\eta,\mu)=\lim_{\Lambda \uparrow
\RR^d}p_\Lambda({\bar \rho}_\Lambda,\eta,\mu)
\ee
and
\be
\frac{\partial p_\Lambda}{\partial \rho}({\bar \rho}_\Lambda,\eta,\mu)=0,
\ee
the lemma follows immediately.
\qed
\par
\blem
\label{L3}
\be
\lim_{\Lambda \uparrow  \RR^d}p_\Lambda(\mu)=\lim_{\eta \to 0 }\lim_{\Lambda \uparrow
\RR^d}p_\Lambda(\eta,\mu).
\ee
\elem
{\bf Proof:}\ \
By Bogoliubov's  convexity inequality one has
\be
-\frac{|\eta|}{\sqrt V}|\<a_0+a_0^*)\>_{H_\Lambda(\mu)}|\leq p_\Lambda(\mu)-p_\Lambda(\eta,\mu)\leq
\frac{|\eta|}{\sqrt V}|\<a_0+a_0^*)\>_{H_\Lambda(\eta,\mu)}|.
\ee
\be
0\leq p_\Lambda(\mu)-p_\Lambda(\eta,\mu)\leq
\frac{2|\eta|}{\sqrt V}|\<a_0^*\>_{H_\Lambda(\eta,\mu)}|\leq
\frac{2|\eta|}{\sqrt V}\<a_0^*a_0\>^\half_{H_\Lambda(\eta,\mu)}
\leq
\frac{2|\eta|}{\sqrt V}\<N_\Lambda\>^\half_{H_\Lambda(\eta,\mu)}.
\label{Q}
\ee
Now, by convexity with respect to $\mu$:
\be
\<\frac{N_\Lambda}{V}\>_{H_\Lambda(\eta,\mu)}\leq p_\Lambda(\eta,\mu+1)-p_\Lambda(\eta,\mu)
\leq p_\Lambda(\eta,\mu+1).
\ee
Since
\bea
H_\Lambda(\eta)-(\mu+1)N_\Lambda
&= &T_\Lambda+\frac{aN^2_\Lambda}{2V}-(\mu+1)N_\Lambda +({\sqrt V}{\bar \eta}+a_0^*)
({\sqrt V}\eta+a_0)-
a_0^*a_0-V|\eta|^2\non\\
&\geq& T_\Lambda+\frac{aN^2_\Lambda}{2V}-(\mu+2)N_\Lambda -V|\eta|^2\non\\
&=& T_\Lambda-(\mu_0-1)N_\Lambda+\frac{a}{2V}\(N_\Lambda-\frac{V(\mu-\mu_0+3)}{a} \)^2\non\\
&& {\hskip 3cm} -V|\eta|^2- \frac{V(\mu-\mu_0+3)^2}{2a}\non\\
&\geq & T_\Lambda-(\mu_0-1)N_\Lambda-V|\eta|^2- \frac{V(\mu-\mu_0+3)^2}{2a},
\eea
we have the estimate
\be
 p_\Lambda(\eta,\mu+1)\leq p_0(\mu_0-1)+1 +|\eta|^2+ \frac{(\mu-\mu_0+3)^2}{2a}
\ee
for large $\Lambda$.
Thus the righthand side of (\ref{Q}) tends to zero as $\eta$ tends to zero.
\qed
\par
Finally as $\eta \to 0$, $\rho(\eta,\mu)$ tends to $(\mu-\mu_0)/a
$ if $\mu \geq \mu_c$ and to the unique solution of $
\rho=\rho_0(\mu-a\rho)$ if $\mu < \mu_c$, proving the Proposition.
\qed
\par
\section{Remarks}
Note that the proof in \cite{D} is based on the equivalence
between the canonical and the grand-canonical ensembles produced
by the mean field interaction, $aN_{\Lambda}^2/2V$, and the use of
explicit information about the grand-canonical free Bose gas.

In the proof presented in \cite{LdeSvdeB} the authors employ the
device of shifting the spectrum to change the density of states
removing the phase transition.

Thirdly, in \cite{DvdeBLP} a probabilistic approach based on the
theory of Large Deviations is used to treat the pressure for this
model, amongst other things.

The essential feature of our proof is the addition of sources to
make the critical density infinite so as to eliminate phase
transitions. This allows us to linearize the mean field
interaction and to control the particle number fluctuations. The
limiting pressure is calculated in this regime and then the phase
transition is restored by the removal of the sources.

Finally we remark that if we replace the interaction term in
(\ref{H}) by the operator $V\phi(N_\Lambda/V)$ where $\phi$ is a
non-negative smooth convex function, the arguments used here go
through with minor modifications to obtain the limiting pressure.

\par
{\bf Acknowledgements:} One of the authors (JVP) wishes to thank the Centre
de Physique Th\'eorique,
CNRS-Luminy  for their kind hospitality and University College Dublin for the award
of a President's Research Fellowship.


\end{document}